\documentclass[pra,aps,floatfix,showpacs,tightenlines,superscriptaddress,amsmath,amssymb,showkeys,10pt,nofootinbib]{revtex4-2}

\usepackage{slashed}
\usepackage{xcolor}
\usepackage{graphicx}

\usepackage{hyperref}

\usepackage{notes2bib}

\newcommand{\be}{\begin{equation}}\newcommand{\ee}{\end{equation}}
\newcommand{\bea}{\begin{eqnarray}}\newcommand{\eea}{\end{eqnarray}}
\newcommand{\brr}{\begin{array}}\newcommand{\err}{\end{array}}
\newcommand{\bit}{\begin{itemize}}\newcommand{\eit}{\end{itemize}}
\newcommand{\ben}{\begin{enumerate}}\newcommand{\een}{\end{enumerate}}

\newcommand{\bbm}{\begin{bmatrix}}\newcommand{\ebm}{\end{bmatrix}}
\newcommand{\ba}{\begin{array}}
\newcommand{\ea}{\end{array}}
\newcommand{\G}{\textbf}

\newtheorem{mydef}{Definition}
\newtheorem{Lemma}{Lemma}
\newcommand{\bd}{\begin{mydef}} \newcommand{\ed}{\end{mydef}}
\newcommand{\bthe}{\begin{theorem}} \newcommand{\ethe}{\end{theorem}}
\newcommand{\ble}{\begin{Lemma}} \newcommand{\ele}{\end{Lemma}}

\newcommand{\dr}{\mathrm{d}}

\def\ha{\frac{1}{2}}

\def\lf{\left}

\def\pa{\partial}

\def\ri{\right}
\def\al{\alpha}\def\Ga{\Gamma}
\def\de{\delta}

\def\Om{\Omega}

\def\1{{_{1}}}\def\2{{_{2}}}

\def\noHe0{:\;\!\!\;\!\!:H_e(0):\;\!\!\;\!\!:}
\def\noHm0{:\;\!\!\;\!\!:H_\mu(0):\;\!\!\;\!\!:}

\def\lf{\left}

\def\pa{\partial}

\def\ri{\right}

\def\al{\alpha}
\def\Ga{\Gamma}\def\de{\delta}

\def\Om{\Omega}

\def\1{{_{1}}}\def\2{{_{2}}}

\begin{document}

\title{Coarsening and metastability of the long-range voter model in three dimensions}

\author{Federico Corberi}
\email{fcorberi@unisa.it}
\affiliation{Dipartimento di Fisica, Universit\`a di Salerno, Via Giovanni Paolo II 132, 84084 Fisciano (SA), Italy}
\affiliation{INFN Sezione di Napoli, Gruppo collegato di Salerno, Italy}

\author{Salvatore dello Russo}
\email{s.dellorusso1@studenti.unisa.it}
\affiliation{Dipartimento di Fisica, Universit\`a di Salerno, Via Giovanni Paolo II 132, 84084 Fisciano (SA), Italy}

\author{Luca Smaldone}
\email{lsmaldone@unisa.it}
\affiliation{Dipartimento di Fisica, Universit\`a di Salerno, Via Giovanni Paolo II 132, 84084 Fisciano (SA), Italy}
\affiliation{INFN Sezione di Napoli, Gruppo collegato di Salerno, Italy}

\begin{abstract}
We study analytically the ordering kinetics and the final metastable states in the three-dimensional long-range voter model where $N$ agents
described by a boolean spin variable $S_i$ can be found in two states (or opinion) $\pm 1$. The kinetics is such that each agent copies the opinion of another at distance $r$ chosen with probability $P(r) \propto r^{-\alpha}$ ($\al >0$). 
In the thermodynamic limit $N\to \infty$ the system approaches a correlated metastable state without consensus, namely without full spin alignment. In such states the equal-time correlation function $C(r)=\langle S_iS_j\rangle$ (where r is the $i-j$ distance) decreases
algebraically in a slow, non-integrable way. Specifically, we find $C(r)\sim r^{-1}$, or $C(r)\sim r^{-(6-\al)}$, or $C(r)\sim r^{-\al}$ for $\al >5$, $3<\al \le 5$ and $0\le \al \le 3$, respectively. In a finite system metastability is escaped after a time of order $N$ and full ordering is eventually achieved.
The dynamics leading to metastability 
is of the coarsening type, with an ever-increasing correlation length $L(t)$ 
(for $N\to \infty$). We find $L(t)\sim 
t^{\frac{1}{2}}$ for $\al >5$, $L(t)\sim 
t^{\frac{5}{2\al}}$ for $4<\al \le 5$, 
and $L(t)\sim t^{\frac{5}{8}}$ for 
$3\le \al \le 4$. For $0\le \al < 3$ 
there is not macroscopic coarsening because stationarity is reached in a microscopic time. Such results allow us to conjecture the behavior of the model for generic spatial dimension. 

\end{abstract}

\maketitle
\section{Introduction}

Albeit Statistical Mechanics is a general approach capable, in principle, to explain the properties of 
basically any physical (and not only) system,
it faces the problem of the overwhelming difficulty of the calculations whenever the
constituents are interacting. Major advances in this field have been promoted by the study of critical systems and phase transitions where 
the concept of universality has been elucidated.
This opened the way to the study of paradigmatic models, such as the Ising one, which,
although perhaps too simple at first sight, 
have been shown to retain the basic ingredients -- the so-called relevant parameters in the renormalization-group language -- characterizing 
a plethora of by far more complicated systems, all belonging to the same universality class. 
However, even with this drastic simplification, 
solvable models are scarce and usually confined to
low dimensional spaces. The aforementioned Ising model, for instance, can only be solved in equilibrium in $D=1,2$~\cite{Ising1925,PhysRev.65.117}.  
As far as non-equilibrium phenomena are investigated, the situation is even worse, and the Glauber solution~\cite{Glauber} of the kinetics in one dimension still remains the only available for this system.
Analytically tractable models, particularly
in $D=3$, are therefore mostly welcome to 
frame in a clear way the physical behaviors. 
In this paper we provide a contribution in this direction by studying the ordering kinetics and the metastablity properties of 
the three-dimensional voter model with long-range algebraic interactions. This article concludes a series of papers~\cite{corberi2023kinetics,corsmal2023ordering,corberi2024aging} where the properties of the model have been addressed in the lower dimensionalities $D=1,2$.

Systems with long-range interactions have been the subject of an increasing interest~\cite{campa2009statistical,book_long_range,DauRufAriWilk}. Perhaps the simplest non-equilibrium process in this context is the phase-ordering kinetics of ferromagnetic materials. These can be effectively modeled by the Ising or related models which, however, 
are not amenable of exact solutions, neither in $D=1$.
This topic has been the subject of many studies~\cite{PhysRevE.49.R27,PhysRevE.50.1900,PhysRevE.99.011301,Corberi_2017,PhysRevE.105.034131,PhysRevE.103.012108,Corberi2021SCI,Corberi2023Chaos,Corberi2023PRE,PhysRevE.102.020102}. In some of these papers~\cite{PhysRevE.49.R27,PhysRevE.50.1900} the problem has been addressed analytically in arbitrary dimension by means of scaling arguments developed for a continuum approach based on a Ginzburg-Landau free energy functional. The predictions obtained therein have only been put to the numerical test in $D=1,2$~\cite{PhysRevE.99.011301,Corberi_2017,PhysRevE.105.034131,PhysRevE.103.012108,Corberi2021SCI,Corberi2023Chaos,Corberi2023PRE,PhysRevE.102.020102} but, to the best of our knowledge, not in the physically more relevant case with $D=3$. Indeed,
numerical simulation of long-range $3D$ systems
are very demanding, particularly due to the strong finite-size effects introduced by the slow decay of correlations produced by the 
extended interactions. Moreover, the analytical approaches mentioned above are only able to access some properties of the ordering kinetics, such as the asymptotic growth law of the coarsening domains, but cannot describe other features, neither any preasymptotic behavior.

The voter model is an alternative playground where the kinetics with long-range interactions can be considered. 
In its original formulation with nearest neighbor (NN) interactions, it was firstly introduced in the study of genetic correlations~ \cite{Kimura1964,1970mathematics}. Later on its basic properties were derived in Refs. \cite{Clifford1973,Holley1975} and widely studied along the years \cite{Clifford1973,Holley1975,liggett2004interacting,Theodore1986,Scheucher1988,PhysRevA.45.1067,Frachebourg1996,Ben1996,PhysRevLett.94.178701,PhysRevE.77.041121,Castellano09,krapivsky2010kinetic},
along with its many variants~\cite{Mobilia2003,Vazquez_2004,MobiliaG2005,Dall'Asta_2007,Mobilia_2007,Stark2008,Castellano_2009,Moretti_2013,Caccioli_2013,HSU2014371,PhysRevE.97.012310,Gastner_2018,Baron2022} designed to
make it suited for
applications to various disciplines \cite{Zillio2005,Antal2006,Ghaffari2012,CARIDI2013216,Gastner_2018,Castellano09,Redner19}.
The model is formed by a collection of $N$ agents described by a spin variable, as in the Ising model. 
However, the elementary dynamical rule is different, in that a randomly selected agent takes the status of one of its NN. In the long-range version of the model considered in this paper, a spin can confront with another one at distance $r$, extracted with probability $P(r)\propto r^{-\al}$, $\al$ being a parameter regulating the decay of interactions. 

The voter model has the advantage of being analytically solvable in any spatial dimension~ \cite{PhysRevA.45.1067,Frachebourg1996} and for any form of the interactions~\cite{corberi2023kinetics,corsmal2023ordering,corberi2024aging}, in the sense that closed equations can be arrived at for
most observables, e.g. (see Eq.~(\ref{eqc1})) for the equal-time correlation function. This allows a precise determination of its behavior
which is not possible in the Ising model. 
It is true that the voter model does not fall into the Ising universality class and, therefore, it cannot be used as a simplified description of magnetic materials. More than that,  detailed balance is violated in this model. However, broadly speaking, it exhibits ferromagnetic behavior and undergoes ordering kinetics characterized by the formation and coarsening of domains, similar to what is observed in other more physically-based magnetic models. 
 
In previous papers~\cite{corberi2023kinetics,corsmal2023ordering,corberi2024aging}, some of us investigated the ordering kinetics of an initially (at time $t=0$) disordered long-range voter model in spatial dimension $D=1,2$. 
In this study, we expand upon that analysis by exploring the evolution of the three-dimensional case. The problem is addressed analytically and the results are benchmarked against numerical solutions. 

Our investigation reveals both similarities and notable differences when compared to lower dimensions. A concise summary of our findings is provided below.
For any value of $\alpha $, the model converges towards a metastable stationary state, with a lifetime that diverges in the thermodynamic limit. 
This is somewhat different to what is found in $D=1,2$~\cite{corberi2023kinetics,corsmal2023ordering} where such metastable states are present only for sufficiently small values of $\al $ ($\al \le 2$ in $D=1$ and $\al \le 4$ in $D=2$).
Such states are only partially ordered,
in the sense that consensus, i.e. full spin alignment, is not present (or, magnetization is zero). The equal-time correlation function $C(r)=\langle S_iS_j\rangle $ exhibits in this condition a slow algebraic decay with the $i-j$ distance $r$, and the correlation length is infinite.
The power-law exponent takes different forms depending on the range of $\al$.
Specifically, for $\al >5$ it is $C(r)\sim r^{-1}$, as in the NN model, while one has $C(r)\sim r^{-(6-\al)}$ for $3<\al \le 5$, and $C(r)\sim r^{-\al}$ for $\al \le 3$.
 
Such stationary states are approached through a coarsening stage  characterized by dynamical scaling and a correlation length increasing as $L(t)\propto t^{\frac{1}{z}}$, with
the growth exponent given by $\frac{1}{z}=\frac{1}{2}$, for $\al >5$,
$\frac{1}{z}=\frac{5}{2\al}$, for $4<\al \le5$, and $\frac{1}{z}=\frac{5}{8}$, for $3<\al \le 4$. For smaller values of $\al$, i.e. $\al < 3$, metastability is attained in a time of order one, and hence there is not a proper coarsening phenomenon.
The lifetime of the metastable state is of order
$N$ for any value of $\al$. After that time, the system quickly attains the fully ordered absorbing state.

Therefore, this paper not only provides one of the rare examples of analytical solutions for systems with long-range interactions far from equilibrium, but it also completes a series of papers aimed at characterizing the features of this model. This study, along with its predecessors, helps to elucidate general characteristics that are likely valid in all dimensions and will be discussed throughout the manuscript.

The structure of the paper is outlined as follows: 
In Section \ref{secmodel}, we introduce the voter model with long-range interactions and derive the evolution equation for the equal-time correlation function, which serves as the fundamental observable for studying most dynamical properties.
In the following Sec.~\ref{lattice} we discuss the expected effects of using different lattices and a possible generalization, dubbed pseudo-lattice, which is particularly suited to solve
numerically the model equations in an efficient way.
Subsequently, Secs.~\ref{secalgt5}, \ref{secalgt3}, and~\ref{secalgt0} focus on the cases with $\alpha > 5$, $3 < \alpha \leq 5$, and $0 \leq \alpha \leq 3$, respectively. Each section provides an in-depth analysis of the model's behavior within the respective parameter range.
The two short sections~\ref{secsizedep} and~\ref{consensus} are devoted to exploring the size dependence of $L(t)$ in the coarsening stage, and of the consensus time $T(N)$, i.e. the time needed to reach the fully ordered absorbing state.
Lastly, in Section~\ref{secconcl}, we recap our findings, compare them with what is observed in lower dimensionalities, present some discussions and draw our conclusions.


\section{The model} \label{secmodel}

The voter model is described by a set of $N$ binary variables located on the nodes $i$ of a graph, which can assume the values $S_i=\pm 1$. In the following we will consider the case of a three-dimensional regular lattice. In the long-range version of the model previously analyzed in Refs.  \cite{corberi2023kinetics,corsmal2023ordering,corberi2024aging} in $D=1,2$, the spins interact with a probability $P(\ell)=\frac{1}{Z}\,\ell^{-\alpha}$, where $\ell$ is the distance between two of them, and $Z$ a normalization.
Distances on a lattice are not, in general, integer
numbers: in the case of a simple cubic lattice in $D=3$, for instance, they are $\ell=1, \sqrt{2}, \sqrt{3}, 2, \ldots$.
In the following we will need to perform summations over all possible distances and, in view of that, we introduce an integer summation index, $p$, indicating the {\it proximity} of two spins at a given distance $\ell$. Namely $p=1$ means NN spins (corresponding on the simple cubic lattice to $\ell=1$), $p=2$ amounts to next-NN ($\ell=\sqrt 2$ on the aforementioned lattice), and so on.
Then 
\be
Z=\sum_pn_p\,\ell_p^{-\al},
\label{eqZ} 
\ee
where $n_p$ is the number of lattice sites at {\it proximity} $p$, and $\ell_p$ the distance between two $p$-neighboring sites. Notice that this notation 
allows us to remain generic with respect to the choice of a particular lattice. In Eq.~(\ref{eqZ}) $p$ runs from 1 to the maximum proximity number in the considered system.
This will be always implicitly understood in any $p$-summation. $n_p$, which is constrained to be a multiple of $2D$, is an irregularly increasing function of $p$, located around an average trend given by
the continuum approximation $n_p=\Omega _{D-1} l_p^{D-1}$, $\Omega _D$ being the surface of a unit $D$-dimensional sphere. Summing up the expression~(\ref{eqZ}), replacing the sum with an integral for large $N$ and assuming the lattice to fill a sphere of radius ${\cal L}\propto N^{\frac{1}{D}}$, one has
\be \label{part}
Z \ \simeq \ \frac{2 \pi^{\frac{D}{2}}}{\Ga\lf[\frac{D}{2}\ri]} \frac{1-{\cal L}^{D-\al}}{\al-D} \, ,
\ee
for $\al \neq D$, and $Z \ \simeq \ \frac{2 \pi^{\frac{D}{2}}}{\Ga\lf[\frac{D}{2}\ri]} \,(\log {\cal L}-1)$, for $\al=D$. $\Ga[x]$ in the equation above is the Euler Gamma function. Notice that $Z$ depends on the number $N$ of spins for $\al \leq D$ also in the thermodynamic limit. 

\begin{figure}[htbp]
	\centering
	\includegraphics[width=0.5\textwidth]{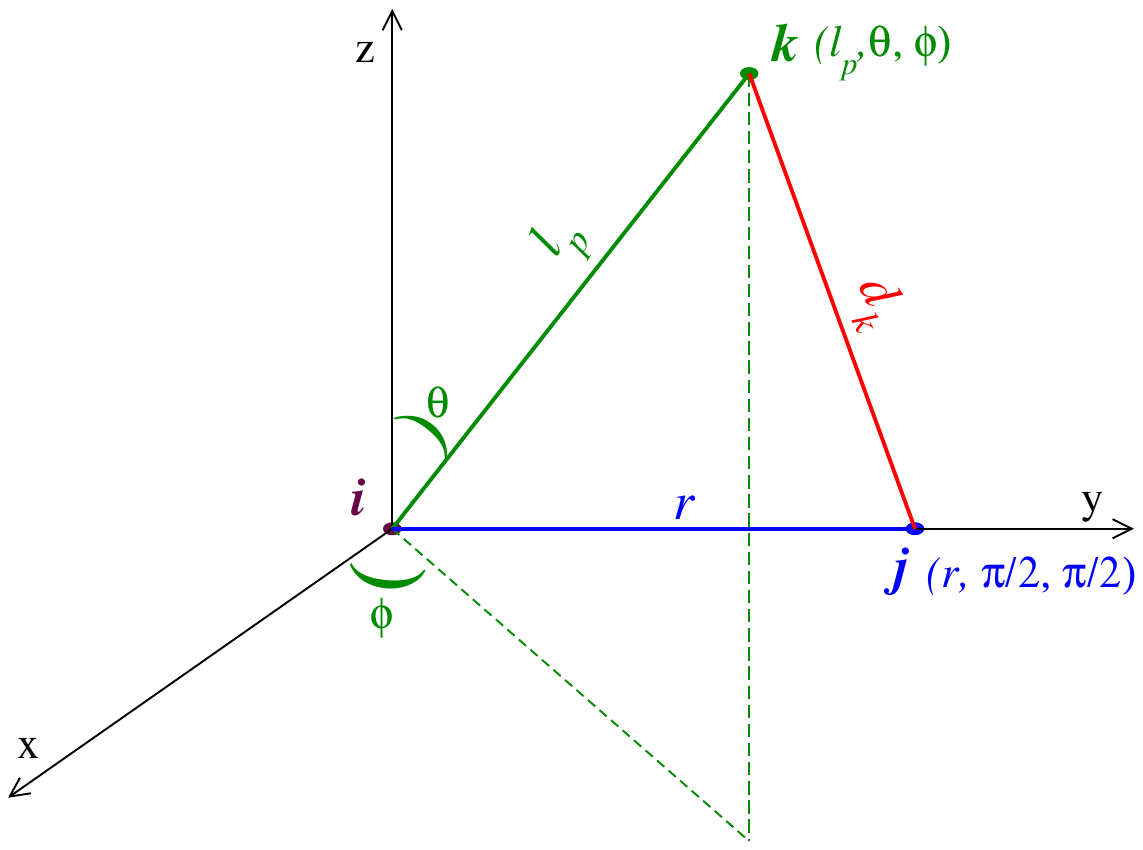}
	\caption{The three points $\boldsymbol i, \boldsymbol j, \boldsymbol k$, are represented in a reference system with origin in $\boldsymbol i$ and $y$-axis on the direction $\boldsymbol i-\boldsymbol j$. The polar coordinates are also indicated.}
	\label{fig:ijkspherical}
\end{figure}

The probability to flip a spin $S_i$ is~\cite{corberi2023kinetics,corberi2024aging,corsmal2023ordering}
\be
w(S_i)=\frac{1}{2N} \, \sum _p P(\ell_p) \sum _{|k-i|=\ell_p}(1-S_iS_k) \, ,
\ee
where $k$ are the $n_p$ sites $p$-neighbors of $i$ and $|k-i|$ means the distance between sites $k$ and $i$.
In this work we will focus on the equal-time correlation functions  $C(r,t) =\langle S_i(t)S_{j}(t)\rangle$, where $r=|i-j|$. Proceeding as in~\cite{corberi2023kinetics,corberi2024aging,corsmal2023ordering},
the evolution equation for this quantity reads
 \begin{eqnarray}
\dot C (r,t) &=&-2 C (r,t) +2\sum _pP(\ell_p) \sum _{k=1}^{n_p}C ([\![d_k(r,\ell_p)]\!],t)\, .
\label{eqc1}
\end{eqnarray}
where the dot indicates the time derivative, as usual. 
The distance $d_k$ is defined as follows: let $\boldsymbol i$, $\boldsymbol j$  and $\boldsymbol k$ be three points on the lattice. In Fig.~\ref{fig:ijkspherical} these are represented on a reference system with origin in $\boldsymbol i$ and $y$-axis directed along the direction $\boldsymbol i - \boldsymbol j$. Of course, there are a certain number $n_p$ of points $\boldsymbol k$ at distance $\ell$ from $\boldsymbol i$, and this is the reason of the sum over $k$ in Eq.~(\ref{eqc1}). With this disposition, $d_k$ is the distance between $\boldsymbol k$ and $\boldsymbol j$. 
The double square-brackets in Eq.~(\ref{eqc1}) mean that we are employing periodic boundary conditions
\be
[\![n]\!]=\left \{ \begin{array}{ll}
n, & \mbox{if }n\in {\cal D} \\
{\cal M}(n), & \mbox{if } n\notin {\cal D},
\end{array} \right .
\ee
where ${\cal D}$ is the set of all possible distances on 1/8 of the lattice, and ${\cal M}(n)$ is the shorter distance computed moving through
the boundary. In a system with long-range interactions like the one we are considering now (for sufficiently small $\al$), observable quantities become size-dependent also 
in the thermodynamic limit. An example is given by the $N$-dependence of the dynamical correlation length $L(t)$ (defined below in Eq.~\eqref{eqL}), reported in Eq.~\eqref{Lsizedepend}. This is at variance with the case of short-range systems. Boundary conditions, therefore, are relevant and must be precisely specified. Those, like open ones, disturbing space translational invariance are not suited for analytical approaches. This is the reason for the choice made here of periodic boundary conditions which, besides these considerations, are also widely used in numerical studies.

The distances $d_k$ can be expressed easily using spherical coordinates. Recalling that the distance $d(\G r,\G r')$ between two points located at $\G r=(r,\theta,\phi)$ and $\G r'=(r',\theta',\phi')$ is
\be
d(\G r,\G r') \ = \ \sqrt{r^2+r'^2-2rr' \lf(\sin \theta \sin \theta' \cos(\phi-\phi')+\cos \theta \cos \theta'\ri)} \, ,
\ee
we get (see Fig.~\ref{fig:ijkspherical})
\be \label{eqd}
d_k(r,\ell _p)=d(j,k) \ = \ \sqrt{r^2+\ell_p^2-2r \ell_p \sin \theta \sin \phi} \, .
\ee

In the following we will study the ordering kinetics of the model after an initial preparation in a fully disordered state, i.e. with $P(S_i)=\frac{1}{2}\delta _{S_i,1}+\frac{1}{2}\delta _{S_i,-1}, \, \forall i$, namely $C(r,t=0)=\delta _{r,0}$, where $\delta $ is the Kronecker function.

\begin{figure}[htbp]
	\centering
		\includegraphics[width=0.8\textwidth]{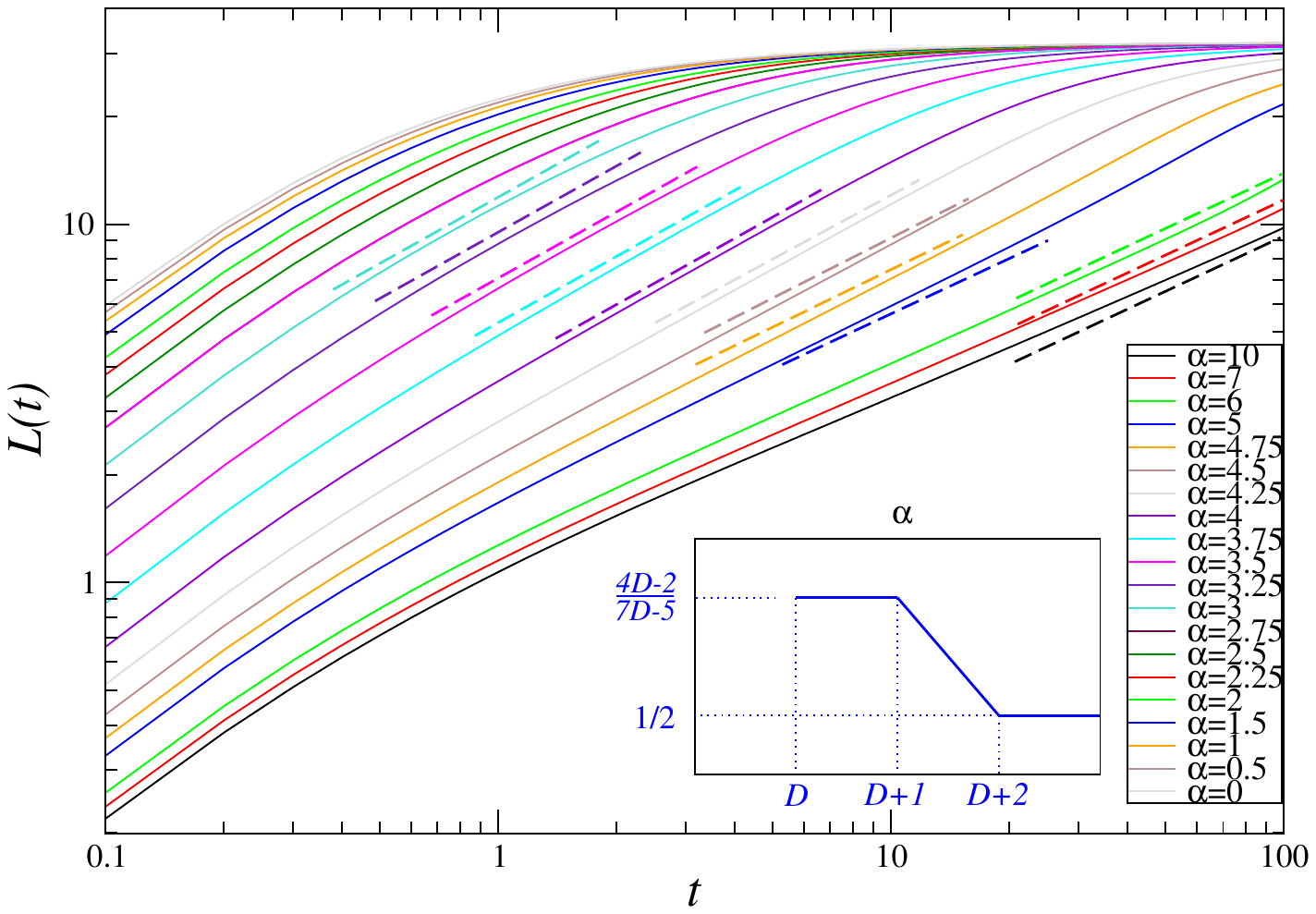}
	\caption{$L(t)$ is plotted against time on double-logarithmic axes, for different values of $\al$, see legend. System size is $N=67^3$. The straight dashed lines are the analytic predictions. The dotted curve is the result obtained on the PL. In the inset, schematic plot of the behavior of the growth exponent $1/z$ ($L(t)\propto t^{1/z}$) as a function of $\al $.}
	\label{fig_length}
\end{figure}

A dynamical correlation length can be extracted from $C$ as 
\be
L(t)=\frac{\sum _{p} n_p \, r_p\,C(r_p,t)}{\sum _{p} n_p C(r_p,t)} \, .
\label{eqL}
\ee
We will see that $L(t)$ may depend on $N$ even asymptotically in some cases (for sufficiently small $\alpha$, see Sec.~\ref{secsizedep}); however we omit such dependence for simplicity. As it can be seen by a preliminary observation of the data in Fig.~\ref{fig_length}, which have been obtained by solving numerically Eq.~(\ref{eqc1}) on a simple cubic lattice and using Eq.~(\ref{eqL}) to extract $L(t)$ from $C(r,t)$, this quantity grows in time while the system orders,
reflecting the presence of larger and larger correlated regions with a certain degree of alignment among the spins. Finally, for large times,
$L(t)$ saturates to a final value $\propto N$
($\gtrsim 30$ in figure) corresponding to a fully ordered state $C(r,t\to \infty)=1$. Taking the thermodynamic limit this saturation limit is pushed further and further and the ordering kinetics lasts 
for longer and longer times (as we will see, this is correct for $\al \ge 3$ only). 
This figure will be discussed further in the following sections. However, one can already appreciate that $L(t)$ increases with different
velocities as $\al $ is varied. 

We remark that, depending on the way it is defined, the average size of domains $L_D(t)$ 
can be
very different from $L(t)$. A possible and frequently used definition of $L_D$ is 
$L_D(t)\propto \rho (t)^{-1}$, where
\begin{equation}
	\rho(t)=\frac{1}{2}[1-C(r=1,t)]
\end{equation} 
is the average interfacial density, i.e. the fraction of antialigned spins. 
As we will show below, while $L(t)$ always grows
unbounded in time (in the thermodynamic limit), 
$\rho(t) $ attains a finite number in any case, due
to the ubiquitous presence of non fully-ordered stationary states. Then $L_D$ does not grow 
at large times. 

For the remainder of this article, we will delve into the model introduced thus far, specifically focusing on the three-dimensional case (but we will also conjecture extrapolations to generic $D$). 

Let us emphasize that the results presented are indeed expected for $t \gg 1$, as will be explained in more detail below. Consequently, our analytical calculations will always be conducted under this condition. The long-time condition is a common assumption in studies of this kind \cite{krapivsky2010kinetic}. Additionally, while it is possible to derive an exact solution that is valid for short time intervals, such a solution would be considerably more complex and less insightful. Approximations were used, in some cases, solely to verify the consistency of the analytical solutions, such as illustrating the behavior of certain integrals. However, these approximations are only for validation purposes and do not affect the accuracy of our solutions. In any case, all the approximations made are confirmed through numerical integration.

\section{Lattice effects} \label{lattice}

The model is fully specified by choosing the form of the lattice. Many dynamical properties, however, are expected to be universal and, hence, independent of such choice. Indeed, in the following sections, most of the analytical calculation will be made by working on a continuum
space, instead of restricting on a specific lattice. This approximation essentially yields exact results for large systems over long timescales. In fact, any corrections would involve powers of $a/L(t)$, where $a$ is the lattice spacing. Since $L(t)$ increases with time, as will be determined in the different $\alpha$ regimes, the continuum approximation should become increasingly accurate for $t \gg 1$. This being said, the definition of the
lattice is unavoidable for an exact solution, in
particular if it is obtained by the numerical integration of Eq.~(\ref{eqc1})~\bibnote{Clearly, the numerical integration of Eq.~(\ref{eqc1}) provides an ``{\it exact}'' solution of the model apart from numerical errors such as truncations, roundings etc...} Looking at Fig.~\ref{fig:ijkspherical} one understands that,
in order to do that, one has to determine the distance $d_k(r,\ell_p)$ between two lattice points 
$k$, $j$, for any possible
choice of $j$ at distance $\ell _p$ from
$i$, i.e. for any value of the angles 
$\theta, \phi$ (provided these angles correspond to lattice points). We did this procedure on a simple cubic lattice and results on this structure will be presented shortly. However this is numerically very time and/or memory consuming, thus severely limiting the values $N$ that can be used. This in some cases hinders the possibility to avoid important finite-size effects.

In order to overcome this difficulty and reach larger values of $N$, enforcing the universality of the dynamical properties mentioned above, we devised an integration scheme of Eq.~(\ref{eqc1})
on an hybrid structure, that we will denote as a 
pseudo-lattice (PL), which incorporates some continuum properties on an underlying lattice. 
Referring again to Fig.~\ref{fig:ijkspherical}, it is defined as follows.
First, we discretize distances in a reasonable yet arbitrary manner. Specifically, we chose the simplest approach where distances take on integer values, similar to a one-dimensional lattice, so that $ r, \ell \in \mathbb{N} $. Second, the angles $ \theta $ and $ \phi $ are assigned $ n_\theta $ and $ n_\phi $ discrete values, respectively, with the constraint $ n_\theta \, n_\phi = \Omega \, \ell^2 $, where $ \Omega $ is a constant. This ensures the 3-dimensional nature of the lattice, meaning the number of points in a shell of radius $ \ell $ is proportional to $ \ell^2 $, with only minor fluctuations due to geometry. In a continuous space, these fluctuations are absent and $ \Omega = 4\pi $, but in general, the value of $ \Omega $ depends on the type of lattice used. On the pseudo-lattice, $ \Omega $ can be treated as a free parameter to optimize performance. We determined $ \Omega = \pi/2 $ and used this value. Additionally, since $ \theta \in [0, 2\pi] $ and $ \phi \in [0, \pi] $, we made the natural choice $ n_\theta = 2n_\phi $ to ensure the same discretization step for both angles.

After defining the allowed values for $ r, \ell, \theta, \phi $, we calculate the distance $ d_k(r,\ell) $ using Eq.~(\ref{eqd}). Since this distance is generally not an integer, it cannot directly serve as the argument of $ C $ in the last term on the right-hand side of Eq.~(\ref{eqc1}). Therefore, we replace $ C([\![d_k(r,\ell_p)]\!],t) $ with a value obtained through interpolation. The simplest approach is to perform a linear interpolation between the values of $ C(r,t) $ at $ r = \text{Int}([\![d_k(r,\ell_p)]\!]) $ and $ r = \text{Int}([\![d_k(r,\ell_p)]\!]) + 1 $, where Int represents the integer part of a number. Since the variations in $ C $ occur on a scale proportional to the growing length $ L(t) $, this straightforward interpolation improves in accuracy over time (although, as we will discuss, in some cases, we observe excellent results even at very short times).

As said, the solution of Eq.~(\ref{eqc1}) on the PL is expected to become equal to the one on any arbitrary $3d$ lattice {\it for sufficiently large times}. Indeed, lattice effects should become negligible because
the length governing the physics is $L(t)$, which 
has grown much larger than the lattice spacing.
We have checked that this is true by comparing the outcomes of the numerical solution of Eq.~(\ref{eqc1}) on the simple cubic lattice and on the PL. An example can be seen in Fig.~\ref{fig_length} where, only for $\alpha=10$ we report the behavior of $L(t)$ computed on the true lattice (continuous black curve) and on the PL (dotted black curve). The two curves are close to each other and tend to coincide for sufficiently long times. 
However, it must be made precise what {\it sufficiently large times} means. For large values of $\al$, say $\al >5$, the evolution of $C(r,t)$ on the cubic lattice and on the PL are practically indistinguishable 
starting from very short times of order $t\simeq 1$. However, upon decreasing $\alpha$, differences appear up to times which become progressively larger. For this reason, for a precise quantitative determination of the growth-law $L(t)$ we preferred to resort to a determination on the cubic lattice (Fig.~\ref{fig_length}), while all the other figures are obtained on the PL, which has the advantage of making larger system sizes accessible.
Indeed, it must be emphasized that, in order to solve Eq.~(\ref{eqc1}) on a regular lattice one has to handle the matrix $d_k(r,\ell_p)$ which is a huge object. For instance, for the case with $N=67$
considered in Fig.~\ref{fig_length} there are 2040
possible distances $r$ (and $\ell_p$) on the square lattice, and $n_p=672$. Then $d_k(r,\ell_p)$ is a 
$2040\times 2040\times 672$ matrix.

Let us also remark that the PL is surely accurate for the study of the stationary states, since these are approached at the largest times. Another advantage of the PL is the fact that $n_p$ is a smooth function of distance, whereas it is a rather irregular function on real lattices. This determines the fact that $C(r,t)$ has some irregular fluctuations when plotted against $r$ if computed on a lattice, while it is a smooth function, much easier to be interpreted, on the PL. This can be appreciated, for instance, in Fig.~\ref{fig_staz_a6}. 	 
	
In the following we study the behavior of the model, analyzing distinct regimes for the parameter $\al$ separately. We emphasize that in the following the pseudo-lattice is used only to determine the stationary states because this is a practical approach to achieve larger system sizes, which would otherwise be unattainable with our available resources. The numerical analysis is primarily intended to validate the analytical results, as it will be seen in the following.
	
\section{Case \boldmath{$\alpha >5$}} \label{secalgt5}

\subsection{Stationary state}

It is known that in the NN case the system tends to a stationary state, where the two-point correlation function has the form \cite{krapivsky2010kinetic},
\be
C_{stat}(r) \ =  \ \frac{a}{r} \qquad \mbox{for }r> a ,
\label{statlarger}
\ee
with $a \ll 1$. We now show that actually this is the case even in the present model, for any $\al >5$.

\begin{figure}[htbp]
	\centering
	\includegraphics[width=0.8\textwidth]{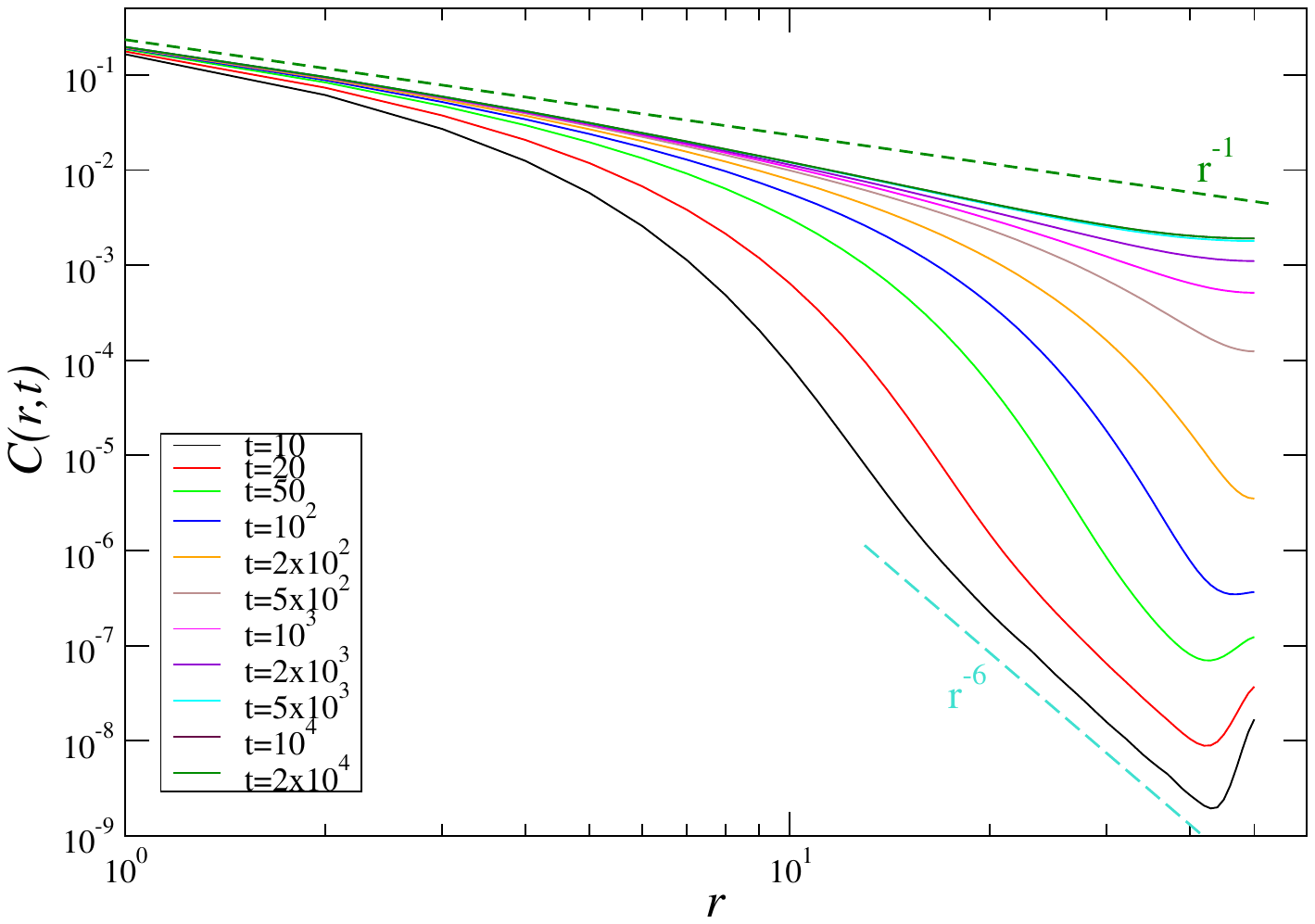}
	\caption{$C(r,t)$, computed on the PL for $\al=6$, is plotted against $r$ on a double logarithmic scale, for different times (see legend). The curves for the last two times are perfectly superimposed. System size is $N=100^3$. The green-dashed line is the analytical expression~(\ref{statlarger}) in the stationary state.  The turquoise long-dashed line is the large-$r$ behavior~(\ref{largexsol1}) during the coarsening stage whereby stationarity is approached. 
	}
	\label{fig_staz_a6}
\end{figure}

Before deriving this result analytically, we can visually confirm that it is correct by looking at Fig.~\ref{fig_staz_a6}, where the evolution of $C(r,t)$ computed numerically by solving Eq.~(\ref{eqc1}) on the PL is plotted for $\al =6$. It is clearly seen that the form~(\ref{statlarger}) (green dashed-line) is approached at small $r$ already at early times (of order $t\simeq 20$) and then propagates to larger and larger values of $r$ as time increases. 
Notice that the curves for the last two times are 
completely superimposed, indicating full stationarization. A slight deviation from the pure
algebraic law~(\ref{statlarger}) is observed at large $r$, which is an effect of the finite size
(increasing $N$ it would be pushed to larger and larger $r$).
Since the system size is finite, the stationary state has a finite life and $C$ eventually starts increasing again. However this occurs on much larger times.   

Now we proceed analytically.
For sufficiently large values of $\al$ we can expand the sum on the r.h.s. of Eq.\eqref{eqc1} in a McLaurin series around $d_k \approx r$, $\Theta \approx \pi/2$, $\Phi \approx \pi/2$, where the capital Greek letters indicate the spherical coordinates in a system centered in $j$, so that a point $k$ has coordinates $(d_k,\Theta,\Phi)$. After doing that, we will retain only the lowest order terms. Such approximation is justified because, for sufficiently large $\al$ (this will be better specified to be $\alpha >5$ by the self-consistency of the computation), $P(\ell)$ decays so fast
that only the smallest values of $\ell_p$ contribute appreciably to the sum on the r.h.s. of Eq.~(\ref{eqc1}). In other words, only interactions among neighbors really count in such case. Since small $\ell_p$ amounts to $d_k \simeq r$,
the truncated McLaurin expansion is actually sufficient to get the exact result. In such case, Eq.\eqref{eqd} gives
\be \label{eqds}
d_k(r,\ell _p) \approx r-\ell_p  \sin \theta \sin \phi \, .
\ee

The zero order contribution from the expansion cancels the first term on the r.h.s. of Eq.\eqref{eqc1}. 
Further, taking into account that 
\be \label{appang}
\Theta \approx \pi/2-(\ell_p/r) \cos \theta \, , \quad \quad \Phi \approx \pi/2- (\ell_p/r) \sin \theta \cos \phi \, ,
\ee
the first order term cancels because
\be
\int \!\! \dr \Om \, \sin \theta \, \sin \phi \ = \ \int \!\! \dr \Om \, \cos \theta  \ = \ \int \!\! \dr \Om \, \sin  \theta \, \cos \phi \ = \ 0 \, ,
\ee
with $\dr \Om \equiv \dr \theta \, \dr \phi \, \sin \theta$ being the elementary solid-angle.
Then we need to compute the second-order term of the expansion. In order to do this, we remind that the Hessian matrix in spherical coordinates reads

\be
H = \begin{bmatrix}
    \frac{\partial^2 f}{\partial r^2} & \frac{1}{r} \frac{\partial^2 f}{\partial r\partial \Theta}-\frac{1}{r^2} \frac{\partial f}{\partial \Theta} & \frac{1}{r \sin\Theta} \frac{\partial^2 f}{\partial \Phi \partial r}-\frac{1}{r^2 \sin\Theta} \frac{\partial f}{\partial \Phi} \\
   \frac{1}{r} \frac{\partial^2 f}{\partial r\partial \Theta}-\frac{1}{r^2} \frac{\partial f}{\partial \Theta} & \frac{1}{r^2} \frac{\partial^2 f}{\partial \Theta^2} + \frac{1}{r} \frac{\partial f}{\partial r} & \frac{1}{r^2 \sin\Theta} \frac{\partial^2 f}{\partial \Theta \partial \Phi} - \frac{\cos\Theta}{r^2\sin^2\Theta} \frac{\partial f}{\partial \Phi} \\
    \frac{1}{r \sin\Theta} \frac{\partial^2 f}{\partial \Phi \partial r}-\frac{1}{r^2 \sin\Theta} \frac{\partial f}{\partial \Phi} & \frac{1}{r^2 \sin\Theta} \frac{\partial^2 f}{\partial \Theta \partial \Phi} - \frac{\cos\Theta}{r^2\sin^2\Theta} \frac{\partial f}{\partial \Phi} & \frac{1}{r^2 \sin^2\Theta} \frac{\partial^2 f}{\partial \Phi^2}+\frac{1}{r}\frac{\partial f}{\partial r} + \frac{\cot\Theta}{r^2} \frac{\partial f}{\partial \Theta}
\end{bmatrix} \, , 
\ee
where we already set $d=r$.
Thus one finds that, up to second order in the McLaurin expansion, Eq.~\eqref{eqc1} reads
\be
\dot C (r,t) \ = \ \int \!\! \dr \ell \, \dr \Om \,  \ell^2 \, P(\ell) \, \lf[\ell^2 \, \sin^2\theta \, \sin^2 \phi \, \frac{\pa^2 C(r)}{\pa r^2} \ + \   \frac{r^2 \lf( \Phi-\frac{\pi}{2}\ri)^2+r^2\lf(\Theta-\frac{\pi}{2}\ri)^2}{r} \, \frac{\pa C(r)}{\pa r}\ri] \, . 
\ee
Using Eq.\eqref{appang} and the trivial identities
\be
\int \!\! \dr \Om \, \sin^2 \theta \, \sin^2 \phi \ = \ \int \!\! \dr \Om \, \sin^2 \theta \, \cos^2 \phi \ = \ \int \!\! \dr \Om \, \cos^2 \theta \ = \ \frac{4 \pi}{3} \, ,
\ee
one easily shows that Eq.~\eqref{eqc1} has the form of a diffusion equation
\be \label{diffueq}
\dot C(r,t)  = \ \mathcal{J} \Delta C(r,t)  \, .
\ee
 Here $\Delta \equiv {\rm Tr} H \ = \  \frac{\pa^2}{\pa r^2}+\frac{2}{r} \frac{\pa}{\pa r}$ is the Laplace operator, while $\mathcal{J} \equiv \frac{4 \pi}{3} \int \dr \ell \, \ell^4 P(\ell)$, which only converges for $\al >  5$. This establishes the limit of validity of the current calculation based on the McLaurin expansion. For smaller values of $\al$ the calculation must be carried over differently, as it will be detailed in Secs.~\ref{secalgt3},\ref{secalgt0}. Computing ${\cal J}$ explicitly one has
\be
{\cal J} \ = \ \frac{\alpha -3}{3 (\alpha -5)} \, .
\label{eqJ}
\ee
This quantity decreases with increasing $\al$ and tends to $1/3=1/D$ when $\al \to \infty$, i.e. in the NN limit. Indeed this is the correct value computed for the NN model in~\cite{krapivsky2010kinetic}. 

If one looks at stationary solutions ($\dot{C}=0$), Eq.\eqref{diffueq} amounts to the Laplace equation
\be
\Delta C(r) \ = \ 0 \, .
\ee
This has to be solved with the boundary condition $C(0)=1$. However, following what is done in the continuous approximation in the NN case \cite{krapivsky2010kinetic}, one can insert a small distance $a$ as regularization and impose $C(r=a)=1$. Therefore, the problem coincides with that of finding the electrostatic potential generated by a sphere of radius $a$, whose boundary is maintained at a unit potential. We thus recover the result \eqref{statlarger}. 

We can also explicitly verify that the above solution (i.e. Eq.~\ref{statlarger}) is consistent by substituting it in Eq.\eqref{eqc1}. One has
\be
C_{stat}(r)={\sum _p}P(\ell_p) \sum _{k=1}^{n_p}C_{stat} ([\![d_k(r,\ell_p)]\!])\, .
\label{autoconsstat}
\ee
Moving to the continuum approximation one has
 \begin{eqnarray}
C_{stat}(r) &\simeq& \int _1 ^{\cal L}\dr \ell \int \!\! \dr \Omega \, \ell^2 \, P(\ell) \, C_{stat}[\![ d(r,\ell, \theta,\phi)]\!]\, .
\label{eqccon}
\end{eqnarray}
where ${\cal L}$ was defined before Eq.~(\ref{part}) and we wrote $d_k(r,\ell_p)=d(r,\ell,\theta,\phi)$ to make all the dependencies explicit. 
As before, the integrals are dominated by the region $\theta \simeq \phi \simeq \pi/2$, $\ell \ll r$, namely $d_k \approx r$. Using this approximation and the
stationary form~(\ref{statlarger}) on the r.h.s. of Eq.~(\ref{eqccon}) one has
 \begin{eqnarray}
  Z \, C_{stat} (r) & = &  4\pi a r^{-1}\int _1^{\cal L}\dr \ell \,\ell^{2-\al} \ \ \simeq \ \ \frac{4 \pi a}{r \, (\al-3)}\, ,
\label{eqcconnn}
\end{eqnarray}
the last equality holding for ${\cal L} \gg 1$. Recalling that, from  Eq.\eqref{part},
 \be
Z \ \simeq \ \frac{4 \pi}{\al-3}\, ,
\label{part3}
\ee 
Eq.~(\ref{eqcconnn}) is fully consistent.

The stationary correlation length can be computed using the definition~\eqref{eqL}. One finds that this diverges in the thermodynamical limit as 
\be 
L_{stat} \ = \ 
\frac{2}{3} \ {\cal L} \propto \frac{2}{3} \ N^{\frac{1}{3}} \, .
\label{Lstatagt5}
\ee

\subsection{Approaching the stationary state}

In order to study the kinetics for $\al>5$, for distances $r$ smaller of a certain $r^*$, which will be later determined, we have to solve the diffusion equation \eqref{diffueq} with the initial conditions $C(r,0)=\de(r)$ and the boundary condition $C(a,t)=1$. 
The basic solution (heat kernel) of Eq.\eqref{diffueq} is
\be
C_0(r,t) \ = \ \frac{e^{-\frac{ r^2}{4 \mathcal{J} t}}}{(4 \pi \mathcal{J} t)^{\frac{3}{2}} } \, .
\ee
This satisfies the initial condition, while the boundary condition is only fulfilled at $t=0$. Following the same procedure that was adopted in the NN case~\cite{Frachebourg1996}, one can then write the complete solution in the form
\be \label{crt}
C(r,t) \ = \ C_0(r,t) + \int^t_{0} \!\! \dr \tau \, G(t-\tau) \, C_0(r,\tau) \,, 
\ee
where the second piece on the r.h.s. can be tuned as to have the boundary condition respected at each time. One can now perform the Laplace transform of both members of Eq.\eqref{crt} and impose the boundary condition at $r=a$:
\be
s^{-1} \ = \ C_0(r=a,s) + G(s) C_0(r=a,s) \, .
\ee
It easy to compute that
\be
C(r,s) \ = \ \frac{e^{-\frac{r \sqrt{s}}{\sqrt{{\cal J}}}}}{4 \pi  {\cal J} r} \, .
\ee
Because we are interested in the large-$t$ case, we consider $s \ll 1$. The Green's function is thus constant in the large-$t$ limit
\be
G \ \approx \ 4 \pi {\cal J} a \, .
\ee
Then we can explicitly compute $C(r,t)$
\be \label{formf}
C(r,t) \ \approx \ \frac{\,a\,}{\,r\,}\,\text{erfc}\left(\frac{r}{2 \sqrt{{\cal J} t} }\right) \, , 
\ee
where $\text{erfc}$ is the complementary error function and we dropped $C_0$ because it is negligible for $t \gg1$. Note that 
\be
 \lim_{t \to \infty} C(r,t) \ = \ \frac{a}{r} \, ,
\ee
 thus recovering the stationary state form \eqref{statlarger}. By means of the definition \eqref{eqL}, we can compute the correlation length
\be
L(t) \ \approx \ \frac{8}{3}\,\sqrt{\frac{{\cal J}}{ \pi}\,} \, t^{\ha} \, .
\label{glawagt5}
\ee
Such behavior is well-observed in Fig.~\ref{fig_length}. Here the curves for $\al=5,6,7,10$, after a microscopic time of order 
$t\simeq 2$, grow algebraically with an exponent very well consistent with $1/2$. Notice that curves for larger values of $\al$ are lower, as expected due to the form of ${\cal J}$ (see Eq.~(\ref{eqJ}) and discussion below) entering the pre-factor of Eq.~(\ref{glawagt5}). 

A comment regarding the effects of finite-size effects is now in order. As discussed in Sec.~\ref{secmodel}, these produce the flattening of the curves at the plateau value $L(t\to \infty)\gtrsim 30$. However, for intermediate values of $\al $, immediately before
such flattening, the same finite size effects produce a faster growth. This can be appreciated particularly for $\al =4.75, 5, 6$ in Fig.~\ref{fig_length}. This explains why for $\al=5$ the curve starts growing faster at times of order $t\gtrsim 30$, shadowing somehow the $t^{1/2}$
growth-law (in addition, it must be also recalled that $\al=5$, being a critical value below which there is a change of behavior (see Sec.~\ref{secalgt3}), is expected to display logarithmic corrections).

Let us notice that the expression~\ref{formf} cannot be put in a standard scaling form of low-temperature coarsening phenomena~\cite{Corberi_2004}
\begin{equation}
	C(r,t)=f\left (\frac{r}{L(t)}\right )
	\label{scaling}
\end{equation}
Instead, defining the scaling variable $x \equiv r/L(t)$, Eq.~(\ref{formf})
can be cast as
\be \label{formfsc}
C(r,t) \ \approx \ L(t)^{-1 }f \left (\frac{r}{L(t)}\right ),
\ee
with $f(x)=ax^{-1}\,\mbox{erfc}\left (\frac{4}{3\sqrt \pi}\,x\right )$, which resembles the scaling form found in critical coarsening~\cite{Corberi_2004} (i.e. after quenching to the critical temperature) where one finds
\be \label{Cfrac}
C(r,t) \ \approx \ L(t)^{-2(D-D_f) }f \left (\frac{r}{L(t)}\right ),
\ee
where $D_f=D-\beta/\nu$ is~\cite{CONIGLIO2000129} the fractal dimension of the critical correlated clusters ($\beta$ and $\nu$ are the usual critical exponents).
Therefore, the form~(\ref{formfsc}) implies that correlated regions are growing fractal, with a fractal dimension $D_f=5/2$.

The solution obtained thus far, based on a Maclaurin expansion for small $\ell$, is only valid for sufficiently small values of $r$.
Indeed, as one can verify looking at Eq.~(\ref{formf}), for small $r$, $C$ varies much less than $P$ which, in turn, decays sufficiently fast as to make relevant only the contributions produced at small $\ell$. However this no longer true for larger distances (we will determine soon how large). 
In this case, the sum on the r.h.s of Eq.~(\ref{eqc1}) is dominated by the contributions at $k \approx j$, i.e. for $d_k \approx 0$, namely $\ell \approx r$ and $\theta \approx \phi \approx \pi/2$. 
Notice that, according to Eq.~(\ref{formf}), the stationary expression~(\ref{statlarger}) is approached initially for small $r$, extending then to larger 
and larger values of $r$ as time goes by. This can clearly be seen in Fig.~\ref{fig_staz_a6}.
Therefore, in the region $d_k\approx 0$ where the 
sum on the r.h.s. of Eq.~(\ref{eqc1}) takes most contributions, we can approximate $C$ with its stationary form~(\ref{statlarger}).
The first term on the r.h.s. of Eq.~(\ref{eqc1}) can be proven to be sub-dominant and can be discarded, so, going again in the continuum, and using $C(r,t) \ \approx \ a/r$ in the integral,
one has
 \begin{equation}
\dot C (r,t) \approx  2\, P(r) \,  \int \!\! \dr \ell \,\dr \Om \, \ell^2 \, \frac{a}{ d(r,\ell, \theta,\phi)} \, .
\label{eqcbx1}
\end{equation}
 The approximation is correct up to $d\lesssim L(t)$, because for larger distances $C$ departs from the stationary form, see Fig.~\ref{fig_staz_a6}, and decays faster (see Eq.\eqref{largexsol1} below) and gives a negligible contribution to the integral.
Since in the above approximation $d(r,\ell, \theta,\phi) \approx r \lf|\pi-2 \phi\ri| \approx \ell \lf|\pi-2 \phi\ri|$, we can integrate 
up to $\ell \approx L(t)$. Evaluating the integral and
retaining the dominant term for large $L(t)$ one has
\be
\dot C(r,t)  \ \propto \ P(r)  L^2(t) \, . 
\ee
Integrating the differential equation yields
\be
C(r,t)  \ \propto \  P(r) \, \, t^2 \, . 
\ee
In such case the correlation function takes a scaling form
\be \label{largexsol1}
C(r,t) \ \propto \ \lf (\frac{r}{\Lambda(t)}\ri)^{-\al} \, ,
\ee
with 
\be
\Lambda (t) \ \propto \  \, t^{2/\al}.
\ee
Notice that for large $r$, $C$ turns from the fast decay of Eq.~(\ref{formf}) to a slower algebraic decrease, as it can be clearly seen in Fig.~\ref{fig_staz_a6}. The typical length also changes from 
$L(t)\propto t^{1/2}$ to the slower one $\Lambda(t)\propto t^{2/\alpha}$. A similar behavior was also observed in $D=1,2$~\cite{corberi2023kinetics,corsmal2023ordering}.

The value of $x$ which separates the two regimes can be determined requiring the matching of the two forms \eqref{formf} and \eqref{largexsol1}. In order to do this, we remark that for $t \gg 1$ we can use the expression \eqref{formf}. We can then use the asymptotic expression of the complementary error function
\cite{abramowitz1965handbook} 
\be \label{e1as}
\text{erfc}(z) \ \approx \ \frac{e^{-z^2}}{\sqrt{\pi} \,z} \, , 
\ee
valid for large $z$.
We thus find
\be \label{xstar}
x^*(t) \ \propto \ \sqrt{(\al-5) \ln t} \, .
\ee
As time increases, the critical value $x^*$ does the same, pushing this behavior to larger and larger distances, i.e. smaller and smaller correlations. Also this can be spotted in Fig.~\ref{fig_staz_a6}. It is worth noting that as the parameter $\alpha$ approaches $\alpha =5^+$, the critical value $x^*$ tends to zero, indicating that the argument leading to Eq.~(\ref{formf}) becomes invalid for $\alpha \le 5$. This result is analogous to the ones obtained for $D=1$ and $D=2$ \cite{corberi2023kinetics,corsmal2023ordering},
where, however, a factor $\al-3$ or $\al-4$, respectively, was found under the square root in Eq.~(\ref{xstar}).
Then, defining $\al_{SR}=D+2$ as the value of $\al$ above which a behavior analogous to the NN case is found, we argue that 
the following expression
\be \label{xstard}
x^*(t) \ \propto \ \sqrt{(\al-\al_{SR}) \ln t} \, .
\ee
could apply to any dimension $D$.

\section{Case $3< \al \leq 5$} \label{secalgt3}
\subsection{Stationary state}
In such case we will show that the stationary expression for the correlation function is 
\bea
C_{stat}(r) \ \propto \ 
      r^{-(6-\al)},  \qquad \quad \mbox{for }r > 1\,. 
   \label{statClt5}
\eea

\begin{figure}[htbp]
	\centering
	\includegraphics[width=0.8\textwidth]{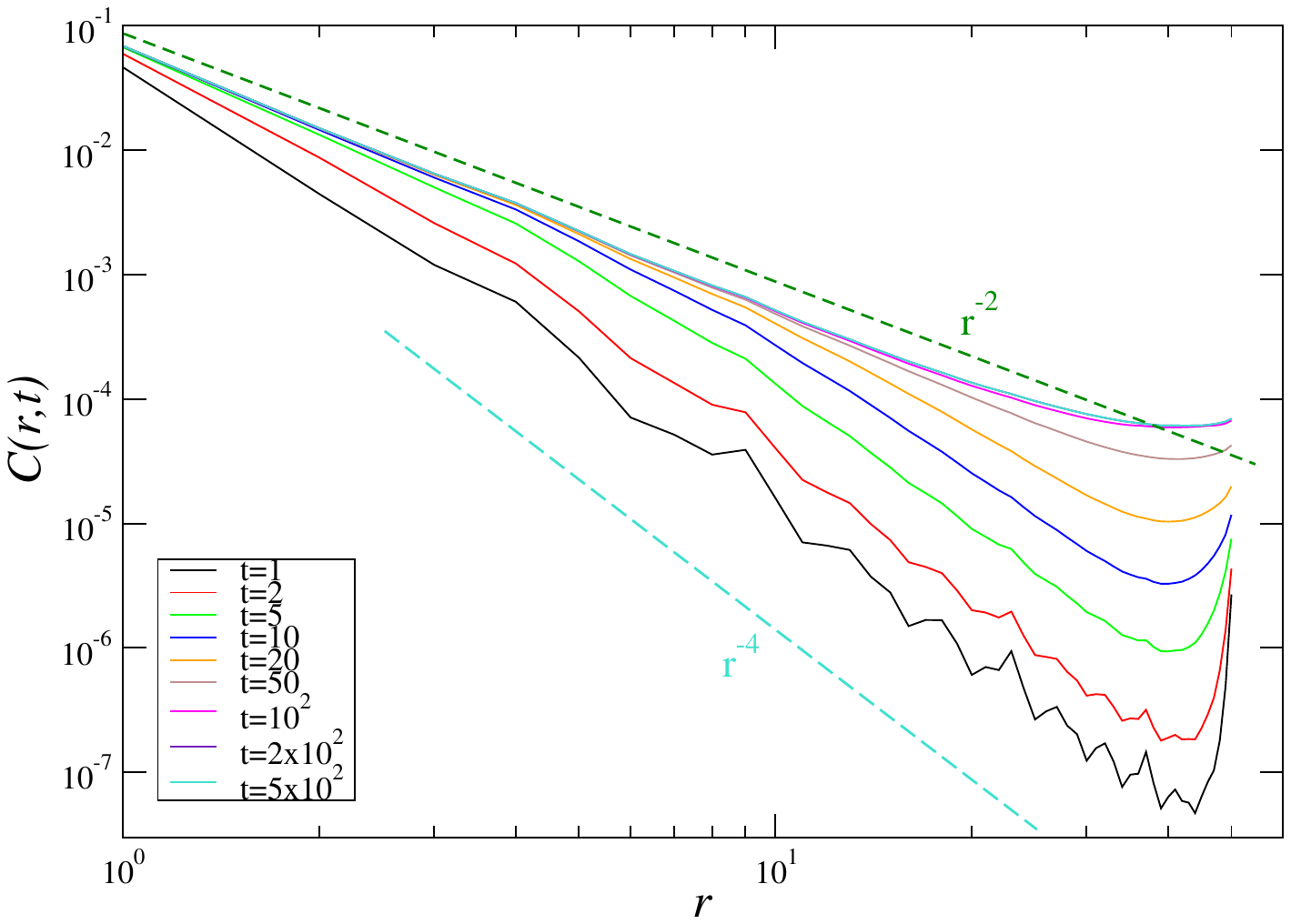}
	\caption{$C(r,t)$, computed on the PL for $\al=4$, is plotted against $r$ on a double logarithmic scale, for different times (see legend). The curves for the last two times are perfectly superimposed. System size is $N=100^3$. The green-dashed line is the analytical expression~(\ref{statClt5}) in the stationary state. The turquoise long-dashed line is the large-$r$ behavior~(\ref{scalfunct}) during the coarsening stage. In the inset the same curves (limited to times $t\le 20$) are plotted against $r/L(t)$.
		}
	\label{fig_staz_a4}
\end{figure}

Postponing the analytical argument, we can verify that this is correct by looking at Fig.~\ref{fig_staz_a4}, showing the evolution of $C(r,t)$ computed numerically by solving Eq.~(\ref{eqc1}) on the PL for $\al =4$. As in the case $\al >5$, the stationary form~(\ref{statClt5}) (green dashed-line) is approached at small $r$ at early times (of order $t\simeq 5$) and then extends to larger distances.  

 In order to verify Eq.~(\ref{statClt5}) analytically, one should consider Eq.\eqref{autoconsstat} (or Eq.\eqref{eqccon}).
Since for $r\gg 1$ Eq.~(\ref{statClt5}) holds, while
$C_{stat}(0)\equiv 1$, we can use the interpolating form
\be
C_{stat}(r)\simeq (1+\kappa r)^{-(6-\al)},
\label{interpstat}
\ee
where $k$ is a constant, in the sum on the r.h.s. of Eq.~(\ref{eqc1}). Plugging into Eq.~(\ref{eqccon}) we get
 \begin{eqnarray}
  Z \,C_{stat} (r) \ \simeq \  \int _1^{\cal L}\dr \ell \int \!\! \dr \Om \, \ell^{2-\al}\left [1+\kappa \, d(r,\ell,\theta,\phi)\right ]^{-(6-\al)}\, , 
\label{eqccon1}
\end{eqnarray}
where we used Eq.~(\ref{eqd}). 
The integrand is still dominated by the small $\ell$ region so $d(r,\ell,\theta,\phi) \approx r$. Then, for $r \gg 1$ one finds
 \bea
  Z \, C_{stat} (r) & = &  (\kappa r)^{-(6-\al)} \int \!\! \dr \ell  \int \!\! \dr \Om \, \, \ell^{2-\al} \ \approx \  \frac{4 \pi \, r^{-(6-\al)}}{\al-3}  \, ,
\label{eqccon2}
\eea
where the last passage holds for $N \gg 1$. Recalling Eq.~(\ref{part3}), this proves the consistency of the ansatz~(\ref{statClt5}).
 
Using the definition~\eqref{eqL}, the correlation length
in the stationary state diverges, for large-$N$, as 
\be 
L_{stat} \ = \ 
\frac{\al-3}{\al-2} \ {\cal L} \propto \frac{\al-3}{\al-2} \ N^{\frac{1}{3}} \, .
\label{Lstatain35}
\ee
Expression~(\ref{eqccon2})
indicates that the present solution 
can only be valid for $\alpha >3$. 
Indeed a different behavior will be found for $\alpha \le 3$ in Sec.~\ref{secalgt0}.
	
\subsection{Approaching the stationary state}
We argue again that $C(r,t)$ initially approaches the stationary form~(\ref{statlarger}) for small values of $r$, and subsequently this behavior extends to encompass larger and larger distances.
This can be observed in Fig.~\ref{fig_staz_a4}.
Notice that, repeating the argument developed around Eq.~(\ref{Cfrac}), this behavior implies that correlated regions grow with a fractal dimension $D_f=\alpha /2$.
 Additionally, we will demonstrate, by verifying the consistency at the conclusion of the current calculation, that a scaling form~(\ref{scaling}) for $C$ 
is obeyed at large $r$, with 
\begin{equation}
	f(x)\propto x^{-\al},\qquad \mbox{for }x\gg 1,
	\label{scalfunct}
\end{equation}
where $L(t)$ has to be computed.
This can be observed in the inset of Fig.~\ref{fig_staz_a4},
	where it is seen that curves for $C(r,t)$ at different times tend to collapse when plotted against $x=r/L(t)$ (where $L(t)$ has been obtained through Eq.~(\ref{eqL})) .
Plugging the scaling assumption~(\ref{scaling}) into Eq.~(\ref{eqc1}) we arrive at
\be 
\frac{\dot{L}(t)}{L(t)} \,  x \, f^{'}(x) \ = \ S(x)\, ,
\label{eqscal}
\ee
with
\be \label{diffeqtrue}
S(x) \ = \ 2 \lf[ f(x) -\sum _p P(\ell_p) \sum _{k=1}^{n_p}f\lf(\frac{[\![d_k(r,\ell_p)]\!]}{L(t)}\ri)\ri] \, .
\ee
Once again we have to evaluate the sum appearing in 
Eq.~(\ref{diffeqtrue}). 
We work for large $r$.
For small $\ell \ll r$ we can set $f([\![d_k]\!]/L(t))\simeq f(x)$ in the sum on the r.h.s. of Eq.~(\ref{diffeqtrue}). The contribution originated in this way cancels with the first term on the r.h.s. What remains is due to the pronounced peak of the correlation at $d_k\simeq r$. This contribution can be estimated as the volume of an iper-solid whose maximal height is $P(r)$ and with a "base" (actually a volume in $3D$) of size 
$V$ which we are not currently able to determine. Putting everything together one has $S\propto r^{-\al}V$. We can argue the following. Firstly, $V$ must be independent of $r$, otherwise we would not have consistency with respect to $x$ once plugged $S$ into Eq.~(\ref{eqscal}). Regarding the $L$ dependence, we expect it to be algebraic as $V\propto L(t)^{3\beta}$, where $\beta $ has to be determined. Then $S\propto r^{-\alpha}L^{3\beta}$. After inserting this value into Eq.~(\ref{eqscal}) the $x$ dependence cancels and, solving the resulting equation for $L$ we arrive at $L(t)\propto t^{\frac{1}{\al -3\beta}}$. 
Now, working for $\al >4$ (the reason will be explained below) 
we make the simplest hypothesis that $\beta $ is a linear function of $\al$, namely $\beta=a\al+b$, where $a$ and $b$ have to be determined. 
Asking for the growth exponent of 
$L(t)$ to match continuously with the value $1/2$ at $\al=5$ we get
$a=(1-b)/5$. In order to determine 
the value of $b$ we resort to numerical simulations. Inspection of
Fig.~\ref{fig_length} shows that 
the value $b=0$, namely
\be
L(t) \ \propto \ t^{\frac{5}{2 \al}} \, ,
\label{glawain45}
\ee
is very well consistent with the numerical data for all the values of $\al$ in the range considered in the figure (i.e. $\al=4,4.25,4.5,4.75,5$).
Let us notice that 
the choice $b=0$ is somehow natural, 
since it means that $\beta \to 1$ as
$\alpha \to 5^-$, meaning that
the usual relation $V\propto L^3$ is recovered when the short range
behavior (i.e. $\al \to 5^-$) is approached.

On the basis of what we know in $D=1$ and $D=2$, we expect the growth-law~(\ref{glawain45}) to break down at $D=4$. Indeed, it is a common feature in $D=1,2$ that the growth exponent increases upon decreasing $\al$ but only down to a certain value
of $\al_{sat} $ of $\al$ given by $\al_{sat}=D+1$. For smaller values of 
$\al$ (but still for $\al > D$, because there is not a real coarsening stage for $\al < D$, see Sec.~\ref{secalgt0}) the growth exponent remains saturated to the
value reached at $\al_{sat}$.
Our numerical data suggest that something similar occurs also in the present $D=3$ case. Indeed, an exponent larger than $5/8$, namely the one predicted by Eq.~(\ref{glawain45}) for $\al=4$,
is never observed. In addition, the exponent $5/8$ is consistent with the behavior of the curves in the coarsening stage.
Defining the growth exponent $1/z$ as $L(t)\propto t^{1/z}$ (in the coarsening stage), this fact promotes its saturation to a maximum value 
\begin{equation}
	\left (\frac{1}{z}\right ) _{max}=\,\frac{4D-2}{7D-5}
	\label{expmax}
\end{equation}
to a general feature which is presumably true in any dimension.

The behavior of the growth exponent $1/z$
is displayed schematically in the inset of Fig.~\ref{fig_length}. This plot summarizes the behaviors found in $D=1,2,3$ and is presumably true for any $D$. The exponent attains its NN value $1/2$ for $\alpha \ge D+2$, it grows linearly up to its maximum value~(\ref{expmax}) 
as $\al$ decreases between $D+2$ and $D+1$ and remains fixed at this value down to $\al =D$. For $\al <D$ there is no coarsening stage (see next section).

\section{Case $0\leq \al  \leq 3$} \label{secalgt0}
We will show below that the following stationary solution
\bea
C_{stat}(r) \ \propto \ 
      r^{-\al},  \qquad \quad \mbox{for }r > 1\, 
   \label{statClt2}
\eea
holds in the small $\al$ regime, i.e. for $\al \leq 3$.

\begin{figure}[htbp]
	\centering
	\includegraphics[width=0.8\textwidth]{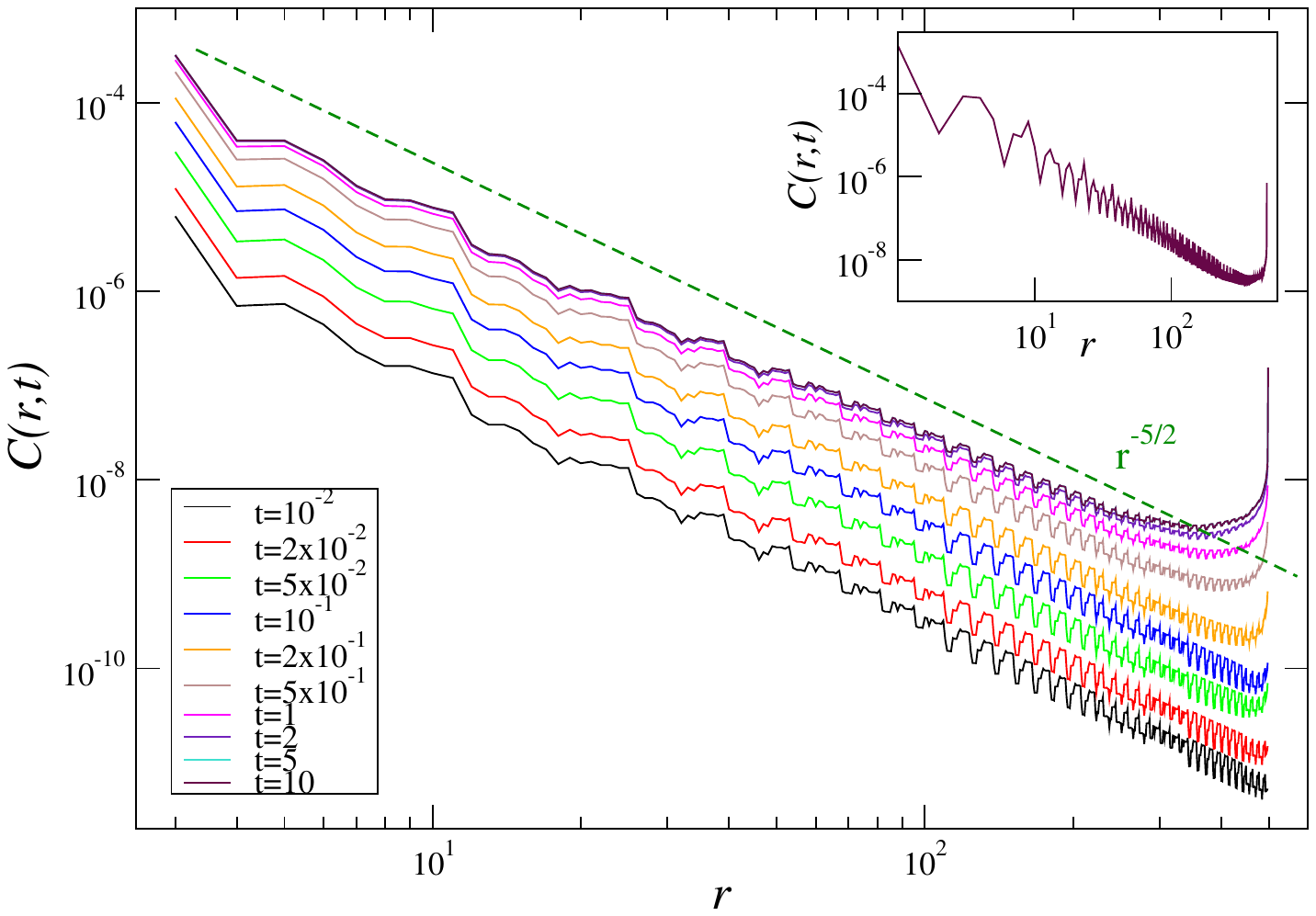}
	\caption{$C(r,t)$, computed on the PL for $\al=5/2$, is plotted against $r$ on a double logarithmic scale, for different times (see legend). In the main panel curves have been smoothed, for better representation, performing a moving average over 5 points. The original curve (for the largest time only) is shown in the inset. The curves for the last two times are perfectly superimposed. System size is $N=1000^3$. The green-dashed line is the analytical expression~(\ref{statClt2}) in the stationary state.
	}
	\label{fig_staz_a2_5}
\end{figure}

This can be appreciated in Fig.~\ref{fig_staz_a2_5}, where the evolution of $C(r,t)$, computed numerically by solving Eq.~(\ref{eqc1}) on the PL for $\al =5/2$, is shown. In this case, the form of $C$ is the one
of Eq.~(\ref{statClt2}) at any time; there is only a time-dependent vertical shift stopping at
stationarity. Notice that, for such small values of $\alpha$, strong lattice effects are present also in the PL, making $C$ rather wiggly (see inset). For this reason in the main part of the figure we plot a
smoothed version of the data obtained by performing a running average over a window containing five points. This makes data easier to be visually interpreted without changing significantly the main features, besides lowering the bouncing. Notice also that the use of the PL allows us to reach sizes as large as $1000^3$.

To assess analytically the validity of Eq.~(\ref{statClt2}) we proceed similarly to what was done in Sec.~\ref{secalgt3}, using the interpolating form
\be
C_{stat}(r)\simeq (1+\kappa r)^{-\al} \, .
\label{interpstat1}
\ee
Putting such expression in the integral in Eq.~(\ref{eqccon}), one has
 \begin{eqnarray}
  Z \,C_{stat} (r) \ \simeq \  \int _1^{\cal L}\dr \ell \int \!\! \dr \Om \, \ell^{2-\al}\left [1+\kappa \, d(r,\ell, \theta,\phi) \right ]^{-\al}\nonumber 
  \, ,
\label{eqccon1bis}
\end{eqnarray}
where we used Eq.~(\ref{eqd}).
The integrand has a peak around $\theta=\phi=\pi/2$, $\ell = r$. The integral is dominated by that region, where we can take $d(r,\ell, \theta,\phi) \approx r \lf|\pi-2 \phi\ri|$. Then, we find
 \bea
  Z \, C_{stat} (r) & \propto &   \frac{{\cal L}^{3-\al} \, r^{-\al}}{3-\al}  \, ,
\label{eqccon3}
\eea
which is consistent with the ansatz \eqref{eqccon} (recall that $Z \simeq 4 \pi\, \frac{{\cal L}^{3-\al}}{(3-\al)}$, see Eq.\eqref{part})).

Let us emphasize that the approximations used to evaluate the integrals in deriving Eq.\eqref{eqccon3} (as well as Eq.\eqref{eqccon2}) and the ones adopted above to evaluate the sum \eqref{diffeqtrue} are intended only as estimations to verify the consistency of the solutions. However, the solutions themselves are expected to be \emph{exact} for $ t \gg 1 $.

Comparing the $D=1$~\cite{corberi2023kinetics}, $D=2$~\cite{corsmal2023ordering} and the present $D=3$ case, we can
try to guess some properties of the stationary states without consensus for general space dimension $D$.
We have already mentioned that a value $\al _{SR}$ exists 
such that, for $\alpha >\alpha _{SR}=D+2$, the behavior of the model is akin to the NN case (as we have seen, this is true both in the aging states and at stationarity).
Going back to partially ordered stationary states, they exist for any value of $\alpha $ (including $\alpha \to \infty$, i.e. NN interactions) only for $D\ge 3$.  Then, for $D\ge 3$ and $\al >\al _{SR}$ stationary states are described by $C_{stat}(r)\propto r^{2-D}$~\cite{krapivsky2010kinetic} (see Eq.~(\ref{statClt5}) for the $D=3$ case studied here), as in the NN case. For $\al \le \al_{SR}$ stationarity without consensus starts to be observed also for $D<3$. More precisely, they appear for $\al \le 4$ in $D=2$ and for $\al \le 2$ in $D=1$.
The form of $C_{stat}$ in this case varies depending on 
$\al _{LR}<\al \le \al _{SR}$ or 
$0\le \al \le \al _{LR}$, where $\al _{LR}=D$ is the characteristic value of $\al $ below which $P(r)$ is non-summable.
For $\al _{LR}<\al \le \al _{SR}$, the form
$C_{stat}\propto r^{-(2 D-\al)}$ is found in $D=1,2,3$~\cite{corberi2023kinetics,corsmal2023ordering} (see Eq.~(\ref{statClt5}) for $D=3$),
so we argue it might be valid for generic $D$. Finally, for $\al \le \al _{LR}$ one has $C_{stat}(r)\propto r^{-\al}$ in $D=1,2,3$~\cite{corberi2023kinetics,corsmal2023ordering} (see Eq.~(\ref{statClt2}) for the present $D=3$ case), which we also guess to be correct for generic $D$. Notice that the various forms of $C_{stat}$ written above and the values of $\al _{SR}, \al _{LR}$ make $C_{stat}$ a 
decreasing non-summable function in any case. Because the decay exponent $C_{stat}(r)\sim r^{-\gamma}$ is related to the fractal dimension of the structure by~\cite{CONIGLIO2000129}
$D_f=D-\frac{\gamma}{2}$ we find that $D_f=\frac{D}{2}+1$, or $D_f=\frac{\al}{2}$, or $D_f=D-\frac{\al}{2}$ in the three cases considered above, namely $\al>\al_{SR}$ (only for $D\ge 3$), $\al_{LR}<\al\le \al_{SR}$, and $0\le \al\le \al_{LR}$, respectively.

Returning to the case $D=3$, using the definition~\eqref{eqL}, the correlation length
in the stationary state diverges, for large-$N$, as 
\be 
L_{stat} \ = \ 
\frac{\al-3}{\al-4} \ {\cal L} \propto \frac{\al-3}{\al-4} \ N^{\frac{1}{3}} \, .
\label{Lstatalt3}
\ee
Notice that for $\al=3$ we have a logarithmic correction $L_{stat} = {\cal L}/\ln {\cal L}\propto 
N^{1/3}/\ln N$.

As in the previous sections, in this case, there exists a macroscopic correlation length at stationarity, indicating that some initial coarsening must characterize the kinetics, as also depicted in Fig.~\ref{fig_length}. However, the time interval over which $L(t)$ increases does not scale with $N$, thus preventing this phenomenon from being observable on a macroscopic scale. This is due to the fact that, as discussed in the following section, $L(t)$ exhibits a dependence on $N$ (for fixed $t$) when $\alpha < 3$. As shown in Eq.~(\ref{Lsizedepend}), indeed, $L(t)$ is of the order of $N^{1/3}$ at fixed $t$, implying that $L(t)$ approaches $L_{\text{stat}}$ in a time of order one, hence microscopic. For this reason, we do not attempt to determine the growth law for $\alpha < 3$. Nevertheless, it is reasonable to expect that the growth exponent remains equal to its maximum value $(1/z)_{\text{max}}=5/8$ obtained at $\al=4^+$, even for $\alpha \le 4$, as suggested by Fig.~\ref{fig_length}. The limited duration of the power-law behavior, however, prevents us from drawing precise conclusions.

\section{Size dependence of the coarsening domains} \label{secsizedep}
We have seen in the previous sections that a scaling form $C(r,t)=f(x) = (a x)^{-\al}$ applies for any value of $\al$, at least at large distances $r$, in the coarsening stage. Using the definition \eqref{eqL}, employing the continuum approximation, one has
\be
\frac{\int^{\frac{{\cal L}}{ L}}_{\frac{1}{L}} \!\! \dr x \,  x^{3} \, f(x)}{\int^{\frac{{\cal L}}{2 L}}_{\frac{1}{L}} \!\! \dr x^2 \,  x \, f(x)} \ = \ 1 \, .
\ee
In the thermodynamic limit what really matters is the form of the integrands for large $x$. Then one finds that, for $\al \neq 4$ and $\al \neq 3$ 
\be
L(t) \ = \ \frac{(\alpha -3) \left({\cal L}^{\alpha }-{\cal L}^4\right)}{(\alpha -4)  \left({\cal L}^{\alpha }-{\cal L}^3\right)}\, . 
\ee
In the limiting case $\al=4$, one has $L \propto \ln {\cal L}$, while for $\al=3$, one gets $L \propto \sqrt {\cal L}/\ln {\cal L}$.  Summarizing
\be
L(t) \ \propto \left \{ \begin{array}{lcl}
\frac{\alpha -3}{\alpha -4}  \, \mathcal{L}^0 \ \propto \ N^0, &\qquad \mbox{for } & \al >4\,, \\[2mm]
\ln \mathcal{L} \ \propto \ \ln N, &\qquad \mbox{for } & \al = 4\,, \\[2mm]
\frac{\alpha -3}{4 -\al}  \, \mathcal{L}^{4-\al} \ \propto \  N^{\frac{4-\al}{3}}, &\qquad \mbox{for } & 3<\al<4\,, \\[2mm]
\mathcal{L}/\ln \mathcal{L} \ \propto \  N^{\frac{1}{3}}/\ln N, &\qquad \mbox{for } & \al =3\,, \\[2mm]
\frac{\alpha -3}{\alpha -4}  \, \mathcal{L} \ \propto \  N^{\frac{1}{3}}, &\qquad \mbox{for } & \al <3 \,.\\[2mm]
\end{array} \right .
\label{Lsizedepend}
\ee

Comparing the cases $D=1$~\cite{corberi2023kinetics}, $D=2$~\cite{corsmal2023ordering} with the present $D=3$ case, one can argue that in general
\be
L(t) \ \propto \left \{ \begin{array}{lcl}
\frac{\alpha -D}{\alpha -D-1}  \, \mathcal{L}^0 \ & \qquad \mbox{for } &  \al >D+1\,, \\[2mm]
\ln \mathcal{L} &\qquad \mbox{for } & \al = D+1\,, \\[2mm]
\frac{\alpha -D}{D+1 -\al}  \, \mathcal{L}^{D+1-\al}  &\qquad \mbox{for } & D<\al<D+1\,, \\[2mm]
\mathcal{L}/\ln \mathcal{L} \  &\qquad \mbox{for } & \al =D\,, \\[2mm]
\frac{\alpha -D}{\alpha -D-1}  \, \mathcal{L} &\qquad \mbox{for } & \al <D \,.\\[2mm]
\end{array} \right .
\ee
\section{Consensus time} \label{consensus}

We now briefly discuss the time needed by a finite system to reach the fully ordered absorbing state -- the consensus time $T(N)$.
This time is composed by the sum of the time 
$T_{coars}$ needed to reach the stationary state
by means of the coarsening dynamics, and the time
to escape it $T_{esc}$, $T=T_{coars}+T_{esc}$. The latter time is $T_{esc}\propto N$ both with short-range interactions~\cite{Krapivsky_Redner_Ben-Naim_2010},
i.e. for $\al=\infty$, 
and in mean-field~\cite{Dben-Abraham_1990}, i.e. for $\al=0$. Then we argue $T_{esc}\propto N, \forall \al$.
Regarding $T_{coars}$, we can assume that
coarsening ends when $L(t)\simeq L_{stat}$.
Recalling the coarsening laws~(\ref{glawagt5},\ref{glawain45}), the $N$-dependence~(\ref{Lsizedepend}), and the forms~(\ref{Lstatagt5},\ref{Lstatain35},\ref{Lstatalt3}) for $L_{stat}$, we get
\be
T_{coars}(N) \ \propto \left \{ \begin{array}{lcl}
	N^{\frac{2}{3}} , &\qquad \mbox{for } & \al >5\,, \\[2mm]
	N^{\frac{2\al}{15}} &\qquad \mbox{for } & 4<\al \le 5\,, 
	\\[2mm]
  N^{\frac{8 (\al-3)}{15}}, &\qquad \mbox{for } & 3<\al<4\,, \\[2mm]
	N^0, &\qquad \mbox{for } & \al <3 \,,\\[2mm]
\end{array} \right.  
\label{Tcoars}
\ee 
 with logarithmic corrections at the boundary values of $\al$. Then $T_{coars}$ is negligible with respect 
to $T_{esc}$, for any $\alpha$, and one has
\begin{equation}
	T(N)\propto N,\qquad \forall \al.
\end{equation}
This is expected also for any $D\ge 3$.

\section{Conclusions} \label{secconcl}

In this paper, we conducted an analytical study of the ordering kinetics of the $D=3$ voter model with long-range interactions, where agents at a distance $r$ agree with a probability distribution $P(r)\propto r^{-\alpha}$. The evolution approaches a non-trivial stationary state with a slow algebraically decaying correlation and a fractal geometry for any value of $\alpha$. Such state is escaped in a time of order $N$, whereafter the absorbing state with full spin alignment is reached. Stationarity, therefore, becomes stable in the thermodynamic limit. Both the form of the stationary state and the 
kinetics leading to it are contingent upon the value of $\alpha$.
For $\alpha > 5$ both the stationary state and the coarsening dynamics preceding it mirror the behavior observed in the nearest-neighbor case with the only difference of a different scaling behavior at large distances, for $x>x^*$ (see discussion starting below Eq.~(\ref{Cfrac})), slowly disappearing as time elapses.  

Instead, for $3 < \alpha \leq 5$, 
the stationary state and the kinetic properties differ markedly from the NN case and become $\al$-dependent. In particular, the presence of a a non-trivial {\it maximum} growth exponent $\frac{1}{z}=5/8$ is found for 
$\al \le 4$. 

Finally, for $\al \le 3$, where $P(r)$ is non-integrable, the stationary state is approached in a microscopic time, as in mean field ($\al =0$).

This paper is the last of a series~\cite{corberi2023kinetics,corsmal2023ordering} where the behavior of the model has been studied in the three physically relevant dimensions $D=1,2,3$. With some differences, notably the presence of stationary states in different ranges of $\al$, many properties are found to be reproduced similarly for all the considered values of $D$. This allowed us to do simple extrapolation conjecture for generic $D$.

The voter model is interesting under many respects. Firstly, on the theoretical side, it is perhaps the only model with a non-trivial ordering kinetics whose solution can be written in closed form (e.g. Eq.~(\ref{eqc1}) in the case considered in this paper) and can therefore be studied in a fully analytical and, in principle, exact way. Secondly, it is a prototypical model underlying many quantitative approaches in many diverse field of knowledge \cite{Kimura1964,1970mathematics,Mobilia2003,Vazquez_2004,MobiliaG2005,Dall'Asta_2007,Mobilia_2007,Stark2008,Castellano_2009,Moretti_2013,Caccioli_2013,HSU2014371,PhysRevE.97.012310,Gastner_2018,Baron2022,Zillio2005,Antal2006,Ghaffari2012,CARIDI2013216,Gastner_2018,Castellano09,Redner19}, ranging from biology, to social sciences and others. Indeed, models frequently used in these contexts can be often considered as variants of the voter one or, at least, regarded as being informed by it. Thirdly, it can be regarded as a proxy to more complicated physical systems as, for instance, magnetic materials as described, for instance, by the Ising or related models. Although the voter model belongs to a different universality class, some of the features enlightened in this and the previous papers~\cite{corberi2023kinetics,corsmal2023ordering} have a clear counterpart in ferromagnetic models. Let us mention here, for instance, the presence of the {\it critical} values $\al_{SR}$ and $\al_{LR}$ separating different regimes where the system falls into the short-range universality class ($\al >\al_{SR}$) or where a fully $\al$-dependent dynamics is observed  
($\al_{LR}<\al\le \al_{SR}$), or when some mean-field features appear ($\al \le \al_{LR}$). Similar features are observed in ferromagnetic models~\cite{PhysRevE.49.R27,PhysRevE.50.1900,PhysRevE.99.011301,Corberi_2017,PhysRevE.105.034131,PhysRevE.103.012108,Corberi2021SCI,Corberi2023Chaos,Corberi2023PRE,PhysRevE.102.020102}. On the basis of these considerations, we hope our results contribute to elucidate the ordering kinetics of magnetic systems with long-range interactions, besides adding a piece of knowledge on the voter model itself. 

Another perspective for future research would be, for instance, the study of percolation properties of the model, as it has already been done in the Ising and the voter model with NN interactions \cite{Saberi_2010,PhysRevE.92.042109} and in the Ising model with long-range interactions \cite{PhysRevE.105.034131}.

\section*{Acknowledgments}

F.C. acknowledges financial support by MUR PRIN 2022 PNRR 2022B3WCFS.

\section*{References}

\bibliography{LibraryStat}

\bibliographystyle{apsrev4-2}

\end{document}